\documentclass[sigconf]{acmart}

\AtBeginDocument{%
  }

\copyrightyear{2025}
\acmYear{2025}
\setcopyright{cc}
\setcctype{by}
\acmConference[WWW Companion '25]{Companion Proceedings of the ACM Web Conference 2025}{April 28-May 2, 2025}{Sydney, NSW, Australia}
\acmBooktitle{Companion Proceedings of the ACM Web Conference 2025 (WWW Companion '25), April 28-May 2, 2025, Sydney, NSW, Australia}
\acmDOI{10.1145/3701716.3715239}
\acmISBN{979-8-4007-1331-6/2025/04}

\usepackage{amsmath}
\usepackage{mfirstuc}
\usepackage{tabularx}
\usepackage{multirow}
\usepackage{makecell}
\usepackage{resizegather}
\usepackage{float}
\usepackage{soul}
\usepackage{stringstrings}
\usepackage{algorithm2e}

\newcommand{\interactrank}{InteractRank}
\newcommand{\itemqueryperf}{ItemQueryPerf}

\newcommand{\unifiedlabel}{Unified Pre-Ranking Label}

\newcommand{\hitsatthree}{HITS@3}
\newcommand{\resultrow}[3]{#1 & #3 & #2 \\}

\begin{document}

\title{InteractRank: Personalized Web-Scale Search Pre-Ranking with Cross
Interaction Features}

\author{Sujay Khandagale}
\orcid{0000-0002-7660-7402}
\affiliation{%
  \institution{Pinterest}
  \city{Palo Alto}
  \country{USA}
}
\email{skhandagale@pinterest.com}

\author{Bhawna Juneja}
\orcid{0009-0001-4688-1435}
\authornote{Equal contribution}
\affiliation{%
  \institution{Pinterest}
  \city{Palo Alto}
  \country{USA}
}
\email{bjuneja@pinterest.com}

\author{Prabhat Agarwal}\authornotemark[1]
\orcid{0000-0002-3826-0858}
\affiliation{%
  \institution{Pinterest}
  \city{Palo Alto}
  \country{USA}
}
\email{pagarwal@pinterest.com}

\author{Aditya Subramanian}\authornotemark[1]
\orcid{0009-0001-3355-9623}
\affiliation{%
  \institution{Pinterest}
  \city{Palo Alto}
  \country{USA}
}
\email{adityasubramanian@pinterest.com}

\author{Jaewon Yang}
\orcid{0009-0001-2224-7915}
\affiliation{%
  \institution{Pinterest}
  \city{Palo Alto}
  \country{USA}
}
\email{jaewonyang@pinterest.com}

\author{Yuting Wang}
\orcid{0009-0009-3367-753X}
\affiliation{%
  \institution{Pinterest}
  \city{Palo Alto}
  \country{USA}
}
\email{yutingwang@pinterest.com}

% \affiliation{%

% \affiliation{%

% \affiliation{%

\renewcommand{\shortauthors}{Sujay Khandagale et al.}

\begin{abstract}

Modern search systems use a multi-stage architecture to deliver personalized results efficiently. Key stages include retrieval, pre-ranking, full ranking, and blending, which refine billions of items to top selections. The pre-ranking stage, vital for scoring and filtering hundreds of thousands of items down to a few thousand, typically relies on two tower models due to their computational efficiency, despite often lacking in capturing complex interactions. While query-item cross interaction features are paramount for full ranking, integrating them into pre-ranking models presents efficiency-related challenges. 

In this paper, we introduce InteractRank, a novel two tower pre-ranking model with robust cross interaction features used at Pinterest. By incorporating historical user engagement-based query-item interactions in the scoring function along with the two tower dot product, InteractRank significantly boosts pre-ranking performance with minimal latency and computation costs. In real-world A/B experiments at Pinterest, InteractRank improves the online engagement metric by 6.5\% over a BM25 baseline and by 3.7\% over a vanilla two tower baseline. We also highlight other components of InteractRank, like real-time user-sequence modeling, and analyze their contributions through offline ablation studies. The code for InteractRank is available at \href{https://github.com/pinterest/atg-research/tree/main/InteractRank}{https://github.com/pinterest/atg-research/tree/main/InteractRank}.
\end{abstract}




\maketitle
\section{Introduction}

\begin{figure}[H]
    \includegraphics[width=1\linewidth]{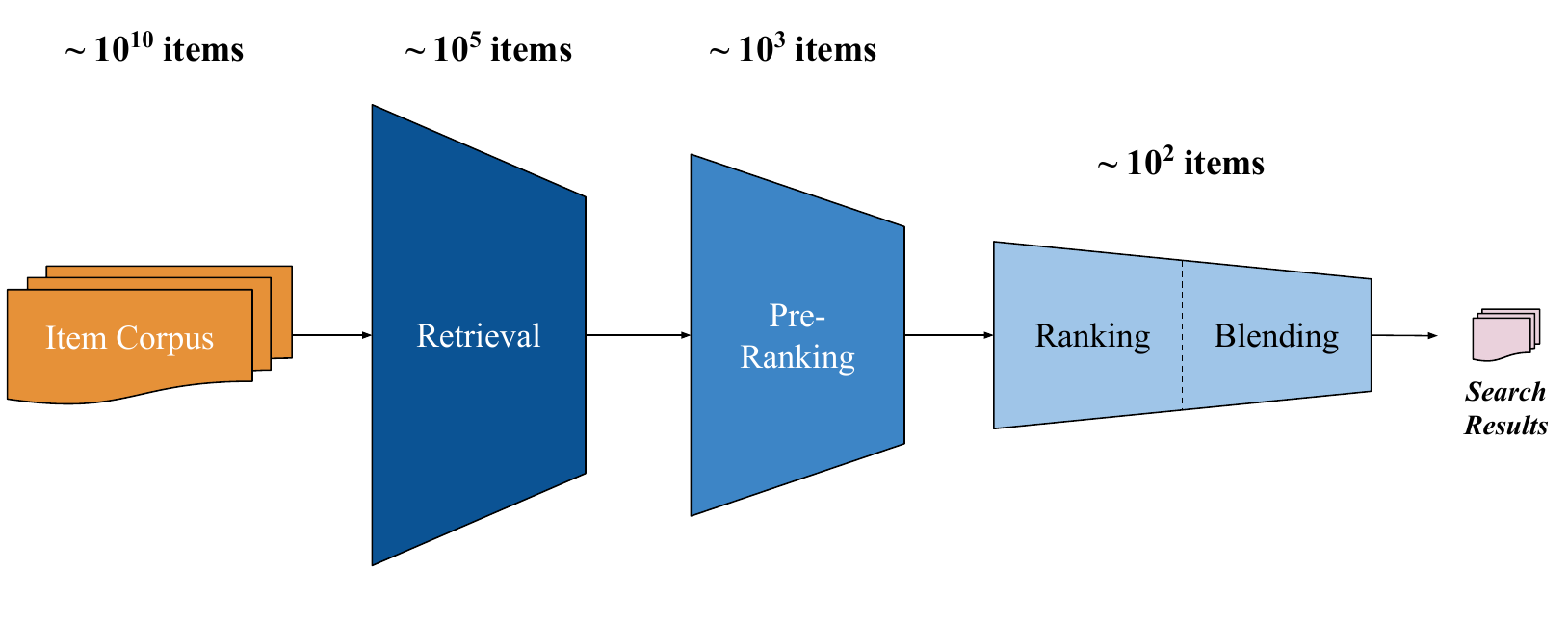}
    \caption{Cascading search recommendation funnel}
    \label{fig:search-funnel}
\end{figure}

Search and Recommendation Systems are a crucial part of any online social media platform. In order to recommend high quality content to users, we need systems that help us score items at a large scale in a short amount of time. These systems typically use a multi-stage pipeline, as shown in Figure~\ref{fig:search-funnel}. The process begins with the retrieval stage, selecting around $10^5$ relevant items from a billion-scale corpus. The pre-ranking stage then uses a lightweight model to refine this list to about $10^3$ items. Following this, the ranking stage employs sophisticated expensive models to narrow down to $10^2$ items. Finally, the blending stage enhances diversity and monetization utility, producing the final items to be shown to the user.

In this paper, we address the challenge of pre-ranking items at a web-scale to deliver substantial product impact. Pre-ranking is essential as it improves the speed and accuracy of search and recommendation systems. By implementing fast pre-ranking methods, we can allocate additional latency budget to run a complex ranking model. Furthermore, pre-ranking plays a vital role in improving the recall, relevance and diversity of the final results.

Developing a pre-ranking method is challenging due to the need to meet strict latency constraints while understanding the complex interactions among users, queries, and items. Given these latency limitations, using heavy deep learning models like DCN and DLRM~\cite{dcnv2, DLRM19} is impractical. A common solution is the two tower model~\cite{10.1145/2959100.2959190}, where one tower computes a query embedding from user and query information, and the other computes an item embedding. Items are ranked based on the similarity between these embeddings. However, while two tower models can capture interactions within a single tower (such as among item features), they fall short in capturing interactions between items and queries, limiting their accuracy.

Another common approach uses count-based features like BM25~\cite{bm25}. These methods create a wide, sparse feature vector based on term occurrences and rank items by aggregating these feature values. While they can capture interactions through co-occurrences, they struggle with capturing the semantic meanings of queries, as they mainly rely on frequency counts rather than conceptual relationships between terms. Consequently, they are also unable to utilize real-time contextual information, such as user activities within the same session.

We introduce InteractRank, a two tower model enhanced with powerful cross-interaction features, designed to combine the strengths of traditional two tower models and interaction-feature-based approaches. Like conventional two tower models, InteractRank constructs a query embedding from user and query attributes and an item embedding from item-specific features, using these to calculate similarity. Beyond this, our model integrates cross-interaction features based on user engagement between items and queries~\cite{click-count}. By combining these cross-interaction features with the dual-tower structure, InteractRank effectively captures item-query interactions and real-time context, such as user activities within the same session.

Our contributions are as follows:
\begin{enumerate}
\item We introduce InteractRank, the production search pre-ranking model at Pinterest serving hundreds of millions of users and billions of searches daily. Through our A/B tests we show that InteractRank significantly boosts user engagement by merging the strengths of the two tower architecture with powerful cross-interaction features. The performance improvement achieved by InteractRank exceeds the sum of the individual contributions of two tower and interaction-feature-based methods, highlighting a strong collaborative effect.
\item We present our framework for generating query-item cross-interaction features, called \itemqueryperf\ (IQP). IQP utilizes item-query relationships derived from two years of user engagement data and has been successfully serving cross-interaction features in our search and recommendation products.

\item We evaluated the impact of various design choices, including feature interaction within each tower, leveraging user’s
activity sequences and the sampling of negative examples,
using large-scale real-world application data. This evaluation provides insights into optimizing two tower models and
cross interaction features for real-world applications.
\end{enumerate}

\subsection{Related Work}
In this section, we review the principal categories of related work relevant to our study: pre-ranking techniques, personalization strategies within recommendation systems, and the utilization of interaction features to enhance model performance.
\subsubsection{Pre-ranking}
Pre-ranking serves as a pivotal component in the modern multi-stage recommendation stack. Its primary function is to narrow down $\mathcal{O}(10k)$ items, sourced from various candidate generators, to a few hundred candidates suitable for complex ranking models~\cite{dcnv2,10.1145/3219819.3220023,10.1145/3219819.3219823,10.1145/3357384.3357925,10.1145/3298689.3347043,10.1145/3383313.3412236,dhen}. Given the substantial volume of items evaluated during the pre-ranking stage, speed is crucial; consequently, two tower models~\cite{dssm} have emerged as a popular choice. Nevertheless, these models are inherently limited in capturing the nuances of interaction between user, query and item. In response, recent studies~\cite{wang2020cold, inttower, santhanam-etal-2022-colbertv2} have aimed to enhance model complexity to improve interaction understanding, while still prioritizing operational efficiency.

There are primarily three types of approaches utilized in enhancing user, query and item interactions. Models such as ColBERT~\cite{colbert,santhanam-etal-2022-colbertv2}, SPLADE~\cite{splade, spladev2}, and SparCode~\cite{sparcode} learn multiple vectors for both queries and items, employing operators like \textit{MaxSim} during serving to balance efficiency with effectiveness. However, many of these methods either incur high deployment costs or require substantial optimization investments, which complicates their integration into existing recommendation stacks.

Another category includes methods like COLD~\cite{wang2020cold} and FCSD~\cite{fcsd}, which adopt architectures that are more akin to those found in ranking models. These approaches tackle the efficiency issue either by designing more efficient architectures or by learning to introduce sparsity into the input features. Despite these improvements, such methods generally remain more costly than two tower models and pose challenges for deployment in web-scale recommendation systems due to significant associated costs.

IntTower~\cite{inttower} represents a hybrid approach between the multi-vector and two tower methodologies. While it learns a dense vector for query and item using InfoNCE loss~\cite{pmlr-v9-gutmann10a}, it promotes early interaction through a MaxSim~\cite{colbert} based score function on the components of the learned vectors and hidden layer representations. Due to the multi-vector approach, this also suffers from issues of increased cost in production.

In contrast to the aforementioned approaches, our method enhances the traditional two tower model by incorporating cost-effective cross-interaction features. This modification is straightforward to integrate into existing systems, enabling deployment at nearly zero additional cost compared to standard two tower models.

\subsubsection{Interaction Features}
User interaction logs provide rich and valuable information for various tasks within a recommendation system. In addition to being used as labels for training multiple models across the recommendation stack, it is commonly used to derive multiple features for ranking models, often leading to substantial performance enhancements~\cite{10.1145/1148170.1148177, 10.1145/2396761.2398691,9416797, 10.1145/3477495.3531948, 10.1145/1277741.1277784}. 
~\citeauthor{10.1145/1031171.1031192}~\cite{10.1145/1031171.1031192} employs the click-document graph to augment the web pages with virtual queries, leading to better relevance learning. ~\citeauthor{10.1145/2911451.2911531}~\cite{10.1145/2911451.2911531} uses label propagation algorithm on the query-document click graph to learn relevance ranking scores for web-pages given a query.
\citeauthor{10.1145/3477495.3531948}~\cite{10.1145/3477495.3531948} provides a framework detailing various methods for extracting features from click logs for learning-to-rank (LTR) models, and it explores the biases associated with these methods. 
Despite their proven benefits, such features derived from user interaction logs have been relatively unexplored in pre-ranking systems due to their inherent complexity and the high costs associated with processing these features at scale. Our work introduces a streamlined approach that efficiently captures the majority of the benefits offered by these features, doing so in a cost-effective manner suitable for large-scale deployment.

\subsubsection{User Behavior History}
Recent trends in recommendation models have increasingly leveraged user behavior history~\cite{10.1145/3326937.3341261,10.1145/3580305.3599918, 10.1145/3219819.3219823, hidasi2015session, 10.1145/3109859.3109900} due to its effectiveness in capturing long-term and evolving user interests, resulting in relevant and personalized recommendations. Typically, methods such as pooling~\cite{10.1145/2959100.2959190, 10.1145/2988450.2988454}, RNNs~\cite{hidasi2015session, 10.1145/3109859.3109877, 10.1609/aaai.v33i01.33015941}, CNNs~\cite{10.1145/3159652.3159656, 10.1145/3109859.3109900}, attention mechanisms~\cite{10.1145/3219819.3219823, 8594844, 10.1145/3336191.3371786}, or transformers~\cite{10.1145/3326937.3341261, 10.1145/3357384.3357895} are employed to automatically learn the intersection of user interests with items. In our work, building upon pooling methods~\cite{10.1145/2959100.2959190}, we propose a learned aggregation of user history to effectively capture user interests.
\section{Methodology}
\begin{figure*}[htb]
    \centering
    % trim=left botm right top
    \includegraphics[clip, scale=0.22, trim={100 80 40 0}]{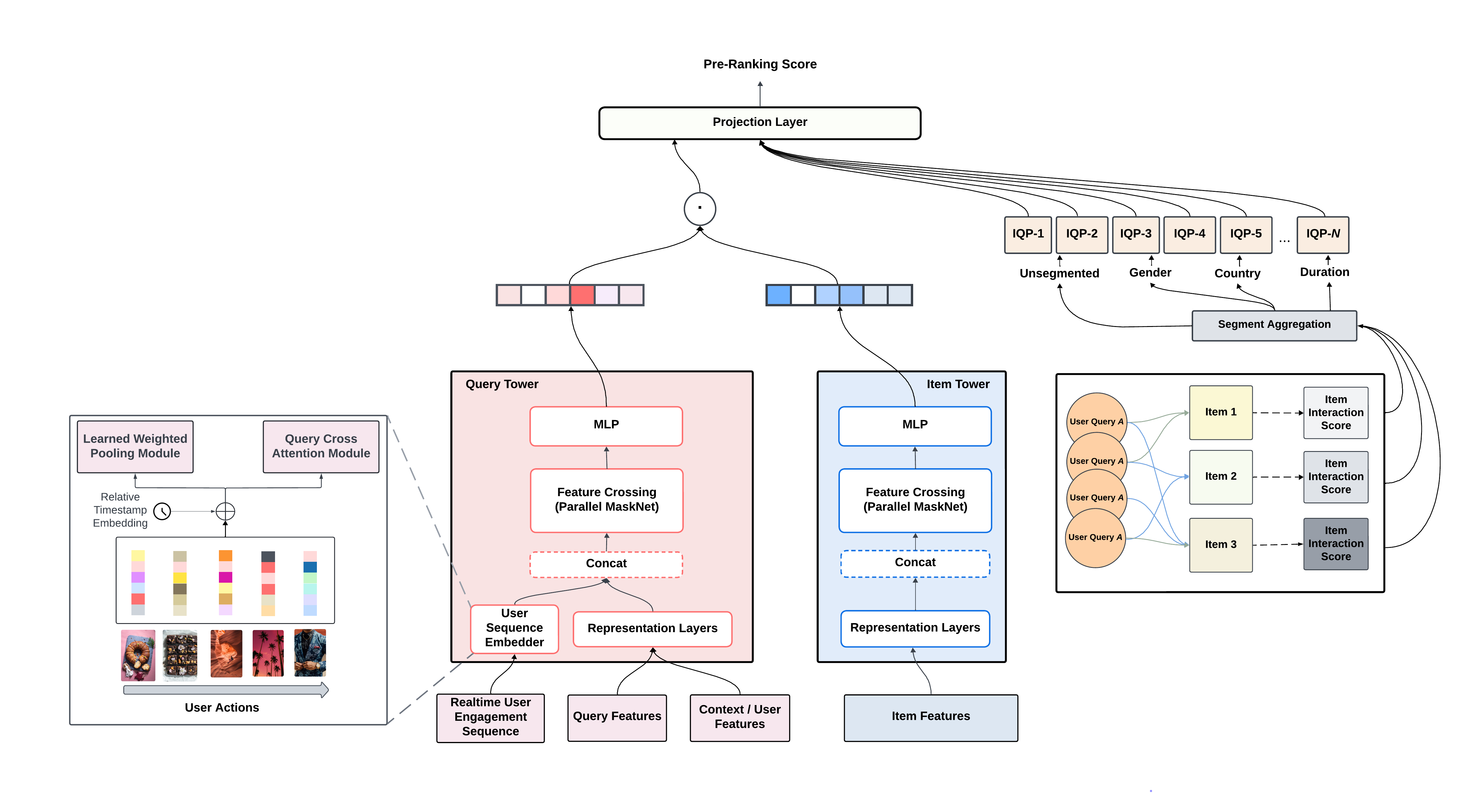}
    \caption{InteractRank Model Architecture Overview. The model embeds the query and item in their towers before combining their dot-product with item-query cross interaction features to generate the final pre-ranking score.}
    \label{fig:model-arch}
\end{figure*}
\subsection{Problem Formulation}
In web-scale search engines, the efficiency and effectiveness of the recommendation stack are key to delivering highly relevant and personalized results. Our system uses a four-stage process—retrieval, pre-ranking, full ranking, and blending—that gradually narrows down the set of items, from billions to hundreds, for each search request. This approach enables the system to handle billions of user requests effectively.

The pre-ranking stage is crucial as it serves as the first filtering phase after retrieval, reducing the vast pool of items to a refined subset for ranking in the next stage. The main challenge for the pre-ranking model is to rank tens of thousands of candidates within mere milliseconds, selecting those that are both relevant to the query and personalized to user preferences. More specifically, given a set of retrieved candidates $\mathcal{I} = {i_1, i_2, ..., i_n}$ and a search request with query $q \in \mathcal{Q}$ and context $c \in \mathcal{C}$ (including user and session data), the pre-ranker learns a ranking function $f: \mathcal{I} \times \mathcal{Q} \times \mathcal{C} \xrightarrow{} \mathbb{R}$ that assigns each candidate a real-valued score, with higher scores indicating better ranks.

In the following section, we provide a detailed description of our proposed method to learn an efficient and accurate pre-ranking model.

\subsection{Two Tower Architecture with Cross Interaction Features}\label{sec:model-archtecture}
We propose \textit{\interactrank}, a two tower model with efficient and inexpensive cross interaction features. As shown in Figure \ref{fig:model-arch}, \interactrank\ jointly embeds the search request and item into a latent embedding space, using the vector dot-product to score item relevance. An affine transformation layer is added after the dot product to enhance this similarity with extra query-item interaction features. This layer captures complex query-item interactions that the dot product might miss, producing a final pre-ranking score that includes both embedding similarity and nuanced query-item cross interactions.

This architecture enables us to: a) maintain efficiency by using decoupled query and item encoders, allowing fleet index lookup for item embeddings at request time, and, b) leverage query-item cross interaction features into our pre-ranking score with minimal impact to serving latency, as features are pre-computed and cached during inference. During training, the two towers and final projection layer are optimized together against a multi-objective loss, as detailed in Section~\ref{sec:training-objective}. Both the query and item towers in \interactrank\ follow a similar structure -- raw features pass through representation layers containing normalization layers and embedding layers, followed by a concatenation layer. This output then goes through in-tower feature crossing and final feed-forward layers to produce the encoded representation.

The ability of two tower models to effectively incorporate multi-modal features has been shown to significantly improve model performance~\cite{huang2020embedding}. For \interactrank, we carefully selected and engineered features to accurately capture the semantics of queries and items, alongside user preferences. In the following sections, we highlight the features consumed by each of the towers.
\subsubsection{Query Tower}
The query tower in \interactrank\ utilizes user representations optimized for historical user engagements~\cite{pancha2022pinnerformer} along with request context features like user metadata (age, gender), geo-location, device type, query language etc. The tower also leverages robust pre-trained query embeddings from OmniSearchSage~\cite{agarwal2024omnisearchsage}, which incorporates image and text features such as titles, descriptions, historically engaged queries, and LLM-generated captions. 

\paragraph{Real-time User Engagement Sequence Modeling} In order to personalize the search results in real-time, 
the InteractRank architecture leverages real-time user engagement sequences in the query tower (as shown in Figure~\ref{fig:model-arch}) alongside user and context features to capture the interests of the user~\cite{xia2023transact}. This approach provides two key benefits: 
1) real-time personalized pre-ranking, with sequences updated in realtime based on current actions, and 
2) personalization beyond search, as the engagement sequence comprises user actions taken across multiple surfaces (e.g. related items, user landing feed). 

Our user engagement sequence comprises a user's 100 most recent engagements, each represented by the engaged item, action type (e.g., save, long click, share), and engagement timestamp. The engaged item is
represented by a diverse set of dense embeddings from different content understanding models. The diversity in the embeddings comes from being trained for different objectives.

The interaction sequence naturally extends itself to sequence modeling architectures. Given the computational constraints of pre-ranking stage, the user sequence embedder in \interactrank\ (in Figure~\ref{fig:model-arch}) comprises two main modules --\\
(a) Learned Weighted Pooling module, which summarizes a user's interests as follows:
\begin{equation}  
    \mathbf{u}_p = \sum_{i=1}^{N} \alpha_i \mathbf{e}_i  
\end{equation}  
where $\mathbf{u}_p$ is the weighted pooled user interest representation, $ N $ denotes the number of interactions in the user's sequence, $ \alpha_i $ represents the learned weight for the $ i^{th} $ interaction, and $ \mathbf{e}_i $ is the embedding of the $ i^{th} $ interaction.  

(b) Query Cross Attention module, which utilizes the query embedding to focus on user engagements relevant to the query: 
\begin{equation}  
    \mathbf{u}_q = \sum_{i=1}^{N} \beta_i(\mathbf{q}) \mathbf{e}_i  
\end{equation}  
  
where $ \mathbf{u}_q $ is the query-focused user representation, $ \mathbf{q} $ is the query embedding, and $ \beta_i(\mathbf{q}) $ is the attention weight for the $ i^{th} $ interaction, calculated using dot-product attention between the query embedding $\mathbf{q}$ and the interaction embedding $\mathbf{e}_i$.  
Both the modules are optimized end-to-end within the model.

\subsubsection{Item Tower}
Within the item tower we incorporate engagement rate and item metadata features. In addition to these, we also include graph-based item embeddings from~\cite{ying2018graph} as well as rich textual embeddings from OmniSearchSage~\cite{agarwal2024omnisearchsage} and image embeddings from~\cite{beal2022billion, zhai2019learning}. 

\subsection{Query-Item Cross Interaction Features} 
User interaction logs provide a wealth of data within recommendation systems, offering deep insights into user preferences and behaviors. However, due to their noisy and large-scale nature, these logs require significant processing to extract meaningful features. In this section, we explore how these noisy, billion-scale query logs can be leveraged to create a set of engagement-based item-query cross-interaction signals known as \itemqueryperf\ (IQP). By utilizing historical search requests and corresponding user engagements, IQP signals are crafted to identify items with a high likelihood of engagement and relevance to a specific query. In essence, we estimate the probability of an item's engagement ($p$) in the context of a query ($q$) using historical logs:
\begin{equation}
    IQP_p(q) = P_{h}(p|q) = \frac{P_{h}(p, q)}{P_{h}(q)} = \frac{C(p, q)}{C(q)}
\end{equation}
Here, $C(p, q)$ indicates how often item $p$ was engaged in search results for query $q$, while $C(q)$ is the total number of times query $q$ appears in the logs. We use smoothing techniques to manage rare queries. Since it's impractical to store the IQP score for all queries for all items, each item's IQP signal includes only the most engaging queries over a specific time period. Consequently, these signals offer a prior engagement and relevance probability for the pre-ranking model.

Logs from multiple time frames—7 days, 90 days, 1 year, and 2 years—are used to estimate these prior probabilities, allowing us to capture insights from both recent and long-term interactions. Additionally, context features like gender, country, and other demographic or behavioral factors can be factored into the prior estimation ($IQP_p(q, c) = P_{h}(p|q, c)$), enabling the IQP model to capture nuanced differences in item interaction among different groups.

These signals must be updated frequently to capture evolving trends, but it's impractical to reprocess two years of logs each time. To address this, we
incrementally update these signals by storing historical counts and revising them with new log data.

\subsection{Training Objective} \label{sec:training-objective}

In a pre-ranking dataset $\mathcal{D}$, each example is represented as a tuple $(q_i, c_i, p_i, y_{a_1}, y_{a_2}, \dots, y_{a_n})$, where $q_i$ is the query, $c_i$ encapsulates user and session context, $p_i$ identifies the item, and $y_{a_k}$ indicates if the user engaged with item $p_i$ via action $a_k$. 

Ranking tasks are typically treated as multi-task issues to capture user interactions' complexity \cite{multi_task_ranking}. The two tower architecture's reliance on vector dot-product scoring in pre-ranking models increases serving costs when training multi-task models that need unique query or item embeddings for each task. To keep the pre-ranking stage lightweight and efficient, we approach it as a single-task problem and optimize our model using a composite label, the \textit{\unifiedlabel}. This label is defined as $U_i = \left(\sum_{a_k \in \mathcal{A}} y_{a_k}\right) > 0$, where $\mathcal{A}$ is the set of actions of interest. It's important to note that not all user actions carry equal significance, and each datapoint can be assigned a weight based on the significance of the actions associated with the item.

Logging candidates at various stages of a multi-stage ranking architecture can be very costly, especially in early stages evaluating tens or hundreds of thousands of items per search request. Due to these constraints, pre-ranking models are often trained on user impression-click data. While useful, this data can lead to distribution mismatch, as it primarily includes engaged items (positives) and those seen but not engaged (implicit negatives), missing ``easy'' negatives - items irrelevant to the user's query.

To address this, we utilize contrastive learning techniques, effective for robust entity representations~\cite{chen2020simple}. This involves generating in-batch negatives by randomly sampling from the mini-batch, allowing the model to better learn by comparing positive examples against various negative examples in the batch.

Hence, we train our model using a weighted combination of binary cross entropy loss with the \unifiedlabel\ and sampled softmax loss with logQ correction~\cite{yi2019sampling, agarwal2024omnisearchsage}.
\begin{equation}
    \mathcal{L} = \phi_e \mathcal{L}_E + \phi_s \mathcal{L}_S.
\end{equation}
Here, the weights $\phi_e, \phi_s$ are the hyperparameters of the model.
The engagement loss $\mathcal{L}_E$ is defined as:
\begin{gather*}
\mathcal{L}_E = -\frac{1}{N} \sum_{i=1}^{N} U_i \log(f(x_i)) + (1 - U_i) \log(1-f(x_i))
% \mathcal{L}_E = -\frac{1}{N} \sum_{i=1}^{N} U_i \log(f(x_i)) + (1 - U_i) \log(1-f(x_i))
\end{gather*}
where, $N$ is the total number of samples, $U_i$ is the \unifiedlabel\ for sample $x_i$, and $f(x_i)$ is the predicted probability that the label is $1$ for sample $x_i$.

The contrastive loss $\mathcal{L}_S$ is defined as:
\begin{gather*}
\mathcal{L}_S = -\frac{1}{|B|} \sum_{i=1}^{|B|} U_i \log \frac{\exp\left(f_c(c_i, q_i) \cdot f_p(p_i)\right)}{\sum_{j=1}^{|B|} \exp\left(f_c(c_i, q_i) \cdot f_p(p_j)\right)}
\end{gather*}
where $f_c$ and $f_p$ represents the query tower and item tower of the pre-ranking model $f$ respectively as described in Section~\ref{sec:model-archtecture}.

As described above, $\mathcal{L}_E$ is focused on the ranking engagement objective, while $\mathcal{L}_S$ addresses the disparities between the training dataset and the actual serving distribution.
\subsection{System Architecture}\label{sys-arch}

The system architecture for pre-ranking includes both offline and online components, as demonstrated in Figure \ref{fig:system-design}. The generation and storage of candidate embeddings are part of the offline workflow, while the online component entails generating query embeddings, conducting a lookup for cross-interaction features, and computing the final pre-ranking score.

\subsubsection{Offline Batch Inference Workflow}

\begin{figure}[htb]
    \includegraphics[clip, scale=0.26, trim={100 100 30 60}]{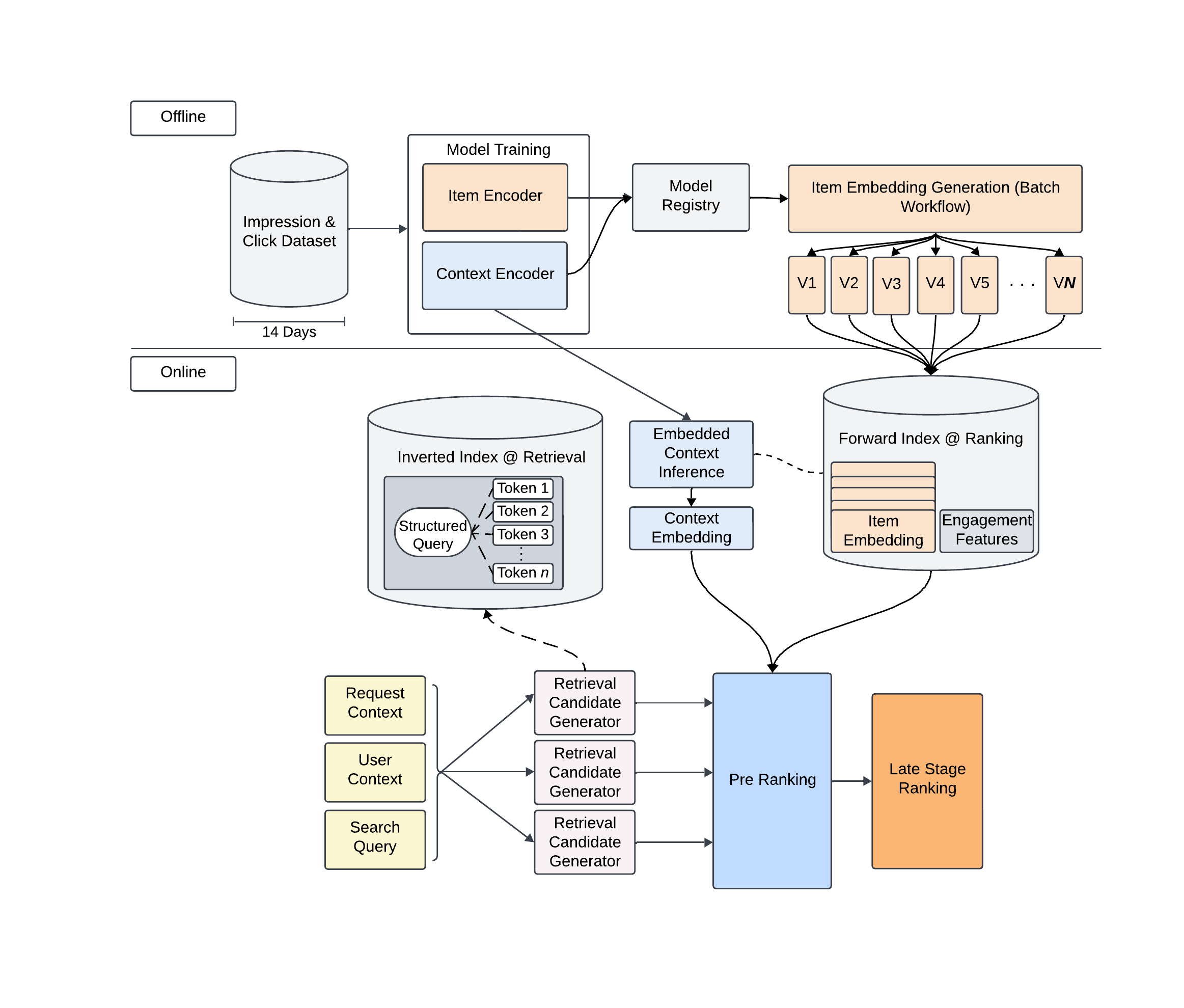}
    \caption{Search pre-ranking system diagram. Training and batch scoring is processed offline, with online retrieval, and multi-stage ranking.}
    % \description{}
    \label{fig:system-design}
\end{figure}

Pinterest corpus comprises billions of items. To enable efficient access to embeddings during online serving, we setup an offline inference workflow which takes in the item features along with the model and generates the corresponding item embedding to a transactional data store. The offline inference workflow runs on a regular cadence to ensure coverage over all available candidates. As seen in Figure \ref{fig:system-design}, the batch inference workflow fetches different types of item features by using 14 days of training data along with model artifacts. All the model inferences happen simultaneously in parallel, which helps in speeding up the workflow runtime. These embeddings are then finally stored in the forward index of the retrieval engine. By storing the embedding in the forward index, it helps to perform a direct lookup during online inference which is an inexpensive operation.

\subsubsection{Online Serving}
Query embedding is generated for each search request using model inference capabilities within the retrieval engine. At a high level, requests can indicate a specific model along with contextual and user information which is passed during the item retrieval stage. An embedded model inference call is made within the retrieval engine to generate the query embedding from the model's query tower. This step is inexpensive and it doesn't add extra overhead to the overall latency of the request. Next, we do an in-memory lookup for the candidate embeddings of the $\mathcal{O}(10^5)$ items matched during retrieval. The corresponding query and item embeddings are used to compute the dot product score. This is then fed as input to a weighted AND (WAND) structured query along with the cross interaction features \& their corresponding weights. Structured query is a tree representation of the original query with node operators like AND, OR or WAND accompanied by weights assigned to each of them. This is ultimately used to compute the final pre-ranking score by implementing InteractRank's final projection layer as structured query operators. 

After executing the pre-ranking logic, the items are sorted based on the pre-ranking score, and then few hundreds of them are sent to final ranking layer for full ranking \& blending.

\subsubsection{Efficient Serving Lookup for Cross Interaction Features}
The IQP signals for a given item-query represent the top K search queries over a time window for which an item received positive engagements like saves, long clicks etc. These features are aggregated offline over different time periods which enables us to capture short term as well as long term user engagement behavior patterns. Since they are stored in forward index, it makes the calculation of the pre-ranking score highly cost-effective. During serving time, we simply perform a lookup in our forward index to obtain these IQP signal values. These are then subsequently added to the final projection layer along with dot product to generate the pre-ranking score.
\section{Experiments and Results}

We evaluate our proposed method both offline and online on Pinterest. Users on the platform can create, browse and search for items. Each item on the platform is composed of media (such as images or videos) and textual descriptions, which may also include hyperlinks to external web pages. 
The users can clickthrough to visit the link associated with the item, save the item to a collection, download or screenshot the associated media, hide/report the item. 
In this section we show the performance improvements from our proposed method through offline ablation studies and online A/B experiments. 
\subsection{Dataset}
We construct our dataset primarily from daily user engagement logs which are generated based on user searches and their corresponding feed views. The logs contain query-item pairs of impressed (i.e shown to the user) items for a search request with their corresponding features and user engagements. The user engagements can be of different types - saves, long clicks, downloads, screenshots, reports, hides.
For training and evaluating our models, we perform a temporal split over a time period $T$. The first $k$ days of data is used for training the model while the last $T-k$ days of data is reserved for evaluating model performance. This method allows us to accurately estimate model performance post-launch in production.
Our platform receives nearly billion daily searches and hence we downsample the search feed views while generating our dataset to keep it tractable and balanced. 
The final dataset statistics can be found in Table \ref{tab:dataset}. 
\begin{table}[bh]
    \centering
    \caption{Summary of dataset statistics. The dataset is constructed from search query logs. M denotes millions, B denotes billions}
    \renewcommand{\arraystretch}{1.25} % Increase the vertical padding 
    \resizebox{\linewidth}{!}{
    \begin{tabular}{c c c c c}
    \hline 
        \textbf{} & \textbf{\# Samples } & \textbf{\# Unique Queries } & \textbf{\# Unique Items } \\
    \hline
        \textbf{Train} & 8.1B & 302.3M & 514.1M \\
        \textbf{Test} & 540M & 32.7M & 117.6M \\
    \hline 
    \end{tabular}
    }
    \label{tab:dataset}
\end{table}
\subsection{Offline Results}
\subsubsection{Offline Evaluation Metrics}

For our offline evaluations, the primary metric used is HITS@K which measures the rank of the items in the top K results after sorting by scores $s$. As seen in the equation below, $rank(s_i)$  is the rank of the items after sorting by predicted score, and $y_i$  is the corresponding label. $\textit{I}$ represents an indicator function which evaluates to 1 when the expression inside evaluates to True and 0 otherwise. 

\begin{equation}
\text{HITS@K}\ ({y}, {s}) = \max\limits_{i|y_i = 1}I[rank(s_i)<=k]
\end{equation}
For our evaluations, we select $K=3$ because it correlates well with online performance.

\subsubsection{Offline Experiment Setup}\label{exp-setup}
All the models in this study are trained for 1.5 epochs on a high-performance compute cluster with 8 NVIDIA A100 Tensor Core GPUs and 320 GB GPU memory. For all the models we train, the encoded embedding size is kept constant at 64 dimensions. This ensures a fair comparison across models by keeping their serving memory footprint constant. During model training, we set the loss weights to $\phi_e = 1.0$ and $\phi_s = 0.01$.
\subsubsection{Comparing InteractRank against other interaction techniques}
In this section, we discuss the offline results from comparing \interactrank\ with other methods. 
We compare our proposed feature interaction method against existing methods. Traditional Two Tower \cite{10.1145/2959100.2959190} architecture relies on the final dot product to capture any cross interactions across the two towers. We consider this method as our baseline since it does not explicitly apply any feature crossing techniques to model for cross-tower interactions. IntTower \cite{inttower} is a recently proposed method that performs explicit fine-grained and early feature interactions to capture the interaction signals between the query and item towers.
Table \ref{tab:intTower} contains the relative improvements of these methods over the baseline. \interactrank\ performs the best, improving \hitsatthree\ on the \unifiedlabel\ as well as long click label both by 2.9\% against the baseline. \interactrank\ also outperforms IntTower on \hitsatthree\ over these labels by 1.5\% and by 0.9\% respectively.
 
The number of serving FLOPs is crucial when serving these models in production in order to control both cost and latency. \interactrank\ achieves these gains over IntTower with nearly 21$\times$ fewer floating point operations (FLOPs) at serving time per search request. Combining both approaches results in even better performance than \interactrank\ alone, indicating the complementary nature of interactions learned from each approach. The combination however still suffers from the significant (24.3$\times$) increase in serving FLOPs, similar to IntTower.

\newcommand{\resultrowneww}[4]{#1 & #2 & #3 & #4 \\}
\begin{table}
\caption{Studying the effects of different feature interaction techniques. We report the \hitsatthree\ metric on the \unifiedlabel\ $(U)$ and long click (LC) binary labels. We also report the number of floating point operations (FLOPs) introduced by each method during serving, assuming a 64-dimensional embedding space. $(^*)$ indicates that the result is from a combination of multiple techniques}
\resizebox{\linewidth}{!}{
\renewcommand{\arraystretch}{1.35} % Increase the vertical padding 
\begin{tabular}{c c c c}
\hline 
\resultrowneww{\textbf{Interaction Method}}{\textbf{\hitsatthree\ $
U$}}{\textbf{\hitsatthree\ LC}}{\textbf{\# FLOPs}}
\hline
\resultrowneww{\makecell{\rule[11pt]{0pt}{0pt}Two Tower \cite{10.1145/2959100.2959190}}}{+0.0\%}{+0.0\%}{127}
\hline
\resultrowneww{\makecell{\rule[11pt]{0pt}{0pt}IntTower \cite{inttower}}}{+1.4\%}{+1.9\%}{3069 (24.2$\times$)}
\hline
\hline
\resultrowneww{InteractRank (Ours)}{\textbf{+2.9\%}}{\textbf{+2.9\%}}{\textbf{142 (1.1$\times$)}}
\hline
\resultrowneww{InteractRank + IntTower$^*$}{\textbf{+4.1\%}$^*$}{\textbf{+3.7\%}$^*$}{3084 (24.3$\times$)$^*$}
\hline
\end{tabular}
}
\label{tab:intTower}
\end{table}

\subsubsection{Ablating components of InteractRank}
In this section, we discuss the impact of remove-one feature ablation on the \interactrank\ architecture to highlight the value of the different components. The results in Table \ref{table:offline-ablation} demonstrate the relative impact of removing the top-3 most impactful model features or architecture components and their impact on HITS@3 for saves and long click labels. Removing user engagement sequence features has the strongest impact on model performance, demonstrating the value from modeling real-time user sequence in \interactrank. Dropping cross interaction features, similarly, causes a heavy degradation in model performance as the model is no longer able to effectively capture cross tower interactions. MaskNet \cite{wang2021masknetintroducingfeaturewisemultiplication} layers within each tower enable them to explicitly learn feature interactions amongst it's features - without these layers, the towers can't learn these in-tower feature interactions, eventually degrading overall model performance. 
\begin{table}[ht]
\caption{Remove-one feature ablation on the InteractRank architecture. We report the relative loss/gain in \hitsatthree\ metric on long clicks and saves. }
\resizebox{\linewidth}{!}{
\renewcommand{\arraystretch}{1.25} % Increase the vertical padding 
\begin{tabular}{c c c}
\hline 
\resultrow{\textbf{Removed Feature}}{\textbf{\hitsatthree\ Saves }}{\textbf{\hitsatthree\ LC}}
\hline  
 \resultrow{User Engagement Sequence}{-13.08\%}{-5.56\%}
 \resultrow{Cross Interaction Features}{-4.47\%}{-3.28\%}
  \resultrow{Parallel MaskNet}{-3.40\%}{-2.29\%}
 
\hline
\end{tabular}
}
\label{table:offline-ablation}
\end{table}

\subsection{Online Results}

% \subsubsection{Experiment Setup}
We conducted an A/B experiment over a large pool of 10M users to evaluate model performance online in production. The experiment spanned a week with statistically significant search traffic where users are randomly picked into the experiment groups.

In the online experiment we compare four models. The baseline model is a standard Proximity BM25 score~\cite{bm25} based pre-ranker which computes the proximity BM25 token match score between the search query and all the item text including title, description, annotations. In addition to this we have 3 treatment models - a) Two Tower model trained with the same optimization objective as described in Section \ref{sec:training-objective}, b) InteractRank model which is described in detail in the previous sections, c) Isolated Cross Interaction Features (ICIF) model which is simply an affine transformation layer over the cross interaction features and optimized for the engagement loss. We strategically choose these experiment groups to isolate and showcase the effectiveness of the InteractRank architecture and its components. 

\subsubsection{Online Metrics}\label{online-metrics}

We evaluate online model performance by directly measuring user satisfaction from their search experience at the session level. 
When a user comes to search on our platform, they have a goal in mind, whether it's something as simple as finding a quick lunch recipe or as complicated as planning a wedding. Therefore, a \textit{search session} can start with a search query and can have multiple subsequent searches and feeds that belong to the same intent. We consider a search session to be \textit{fulfilled} if there is at least one high quality fulfilling action i.e save, long click, download or screenshot in the search session. Our online metric, Search Intent Fulfillment Rate (SIFR) is calculated as,
$$
\text{SIFR} = \frac{\text{\# of fulfilled search sessions}}{\text{\# of search sessions}}
$$
Similarly, we define Fulfilled by 1st Search (F1S) as the fraction of search sessions where a fulfilling action was taken on the first search feed itself. Ideally, we would want users to have their search session fulfilled in as few search queries as possible. Fulfilled by 1st Search helps us to measure for this goal effectively.

\begin{table}[h]
\caption{The relative online improvements of different experiment groups over a Proximity BM25 baseline. $(\uparrow)$ indicates increase in metric is better}
\resizebox{0.9\linewidth}{!}{
\renewcommand{\arraystretch}{1.2} % Increase the vertical padding 
\begin{tabular}{c c c c}
\hline  
\textbf{} & \textbf{SIFR $\uparrow$} & \textbf{F1S $\uparrow$} \\  
\hline  
Two Tower & +2.7\% & +3.9\%\\  
\hline
ICIF & +3.6\% & +5.5\%\\  
\hline \hline
\textbf{\makecell{\rule[0pt]{0pt}{10pt}InteractRank\\(Two Tower + Interaction Features)}} & \textbf{+6.5\%} & \textbf{+9.2\%}\\
\hline
\end{tabular}
}
\label{table:online-results}
\end{table}

\subsubsection{Aggregate Results}
Table~\ref{table:online-results} shows the relative online performance of the 3 treatment models against the proximity BM25 baseline. All three treatments outperform the BM25 baseline on both SIFR and F1S. 
The Two Tower model improves SIFR by 2.7\% and Fulfilled By 1st Search by 3.9\% over the BM25 baseline. This illustrates the importance of utilizing personalized jointly learned entity embeddings for pre-ranking.
We also see that the ICIF model beats Two Tower by a relative improvement of 0.9\% SIFR and 1.6\% F1S. 

By combining both these techniques, \interactrank\ achieves the best performance, improving online SIFR by 6.5\% and F1S by 9.2\% over the BM25 baseline. It also shows strong relative improvement over the Two Tower model, with SIFR and F1S improvements of 3.7\% and 5.1\%, respectively. We also see that the performance improvements from both techniques, i.e a two tower architecture and cross interaction features, are additive. This result underscores the independent and complementary nature of the improvements coming from these approaches, highlighting their combined effect on enhancing the overall model performance.

\subsubsection{Query Popularity Segmented Results}
We broadly categorize search queries into 4 segments (in order) - HEAD, TORSO, TAIL and SINGLE, based on how frequently these are searched over a year.
Table~\ref{table:online-pop-seg-results} highlights the relative performance of the three treatment groups over the BM25 baseline. It is clear that ICIF is better than the Two Tower model only on the HEAD segment. This is expected, as ICIF's estimation of priors based on historical engagements tends not to generalize well to less frequent queries, though it performs strongly for popular ones. By combining the two tower architecture with cross interaction features, \interactrank\ not only aggregates the improvements from these techniques over HEAD queries but also learns a much better tradeoff over TORSO, TAIL and SINGLE queries.
\begin{table}[h]
\caption{Relative improvements in SIFR of treatment models over a Proximity BM25 baseline segmented by query popularity. \textit{Italicized} numbers represent statistically insignificant results}
\resizebox{\linewidth}{!}{
\renewcommand{\arraystretch}{1.25} % Increase the vertical padding 
\begin{tabular}{c c c c c}
\hline  
\textbf{}  & \textbf{HEAD} & \textbf{TORSO} & \textbf{TAIL} & \textbf{SINGLE}  \\  
\hline  
\makecell{Two Tower\\} & +3.4\%	& +2.7\% & +1.7\% & +0.8\% \\  
\hline
ICIF & +5.9\%	& +0.7\% & \textit{-0.1\%} & -0.8\%  \\  
\hline \hline
\textbf{\makecell{\rule[1pt]{0pt}{10pt}InteractRank\\(Two Tower + Interaction Features)}} & \textbf{+8.3\%} & \textbf{+5.0\%}& \textbf{+4.2\%} & \textbf{+1.8\%}  \\
\hline
\end{tabular}
}
\label{table:online-pop-seg-results}
\end{table}

\section{Conclusion}
In this work, we propose a new approach at introducing cross interaction features in two tower models for search pre-ranking. 
Our proposed method, \interactrank, jointly learns a projection layer over a two tower architecture and powerful cross interaction features thereby capturing the cross tower interactions beyond the dot product. 
In contrast to previous works which employ complex architectures to implicitly learn cross tower interactions, we propose the adoption of inexpensive user engagement-based features to model these cross interactions explicitly.
In addition to the cross interaction features, we also highlight other components of our two tower architecture like real-time user sequence modeling which strongly contribute to the model performance. 
Finally, we outline the advantages of our methodology through offline ablation studies and an online A/B experiment. \interactrank\ improves online engagement globally by +6.5\% over a BM25 baseline and by +3.7\% over a two tower baseline. \interactrank\ is currently deployed in production at Pinterest globally, serving billions of search requests daily. 
\section{Acknowledgements}
We would like to thank Krishna Kamath, Kurchi Subhra Hazra, Jiajing Xu, Cosmin Negruseri and Liang Zhang who provided feedback or supported us throughout this project.
% \begin{acks}
%     cdmckdskvs
% \end{acks}
\newpage
\bibliographystyle{ACM-Reference-Format}
\balance
\bibliography{references}

\end{document}